\documentclass[12pt,preprint]{aastex}
\usepackage{psfig}
\def\HI {H\kern0.1em{\sc i}} 
\def\deg{$^{\circ}$}

\begin{document}
\title{~~\\ ~~\\ Dating COINS: Kinematic Ages for Compact Symmetric Objects}
\shorttitle{CSO Motions}
\shortauthors{Gugliucci et al.}
\author{N.E. Gugliucci\altaffilmark{1,2}, G.B. Taylor\altaffilmark{2,3},
  A.B. Peck\altaffilmark{4}, M. Giroletti\altaffilmark{5}}
%\affil{}
\email{gugnico@lycoming.edu; gtaylor@nrao.edu; apeck@cfa.harvard.edu;
  giroletti@ira.cnr.it}
\altaffiltext{1}{Lycoming College, Williamsport, PA 17701}
\altaffiltext{2}{National Radio Astronomy Observatory, P.O. Box O, Socorro, NM 87801}
\altaffiltext{3}{Kavli Institute of Particle Astrophysics and Cosmology,
Menlo Park, CA 94025}
\altaffiltext{4}{Harvard-Smithsonian CfA, SAO/SMA Project, 645 N. A'ohoku Pl., Hilo, HI 96720}
\altaffiltext{5}{Istituto di Radioastronomia del CNR, via Gobetti 101,
  40129 Bologna, Italy}
%\received{}
%\accepted{}
%\journalid{}{}
%\articleid{}{}

\slugcomment{As Accepted to ApJ}

\begin{abstract}

We present multi-epoch VLBA observations of Compact Symmetric Objects
(CSOs) from the COINS sample (CSOs Observed In the Northern Sky).
These observations allow us to make estimates of, or place limits on, the kinematic ages for those sources with well-identified hot spots.  This study
significantly increases the number of CSOs with well-determined ages
or limits.  The age distribution is found to be sharply peaked under
500 years, suggesting that many CSOs die young, or are episodic in
nature, and very few survive to evolve into FR~II sources like Cygnus~A.
Jet components are found to have higher velocities than hot spots
which is consistent with their movement down cleared channels. We also
report on the first detections of significant polarization in two
CSOs, J0000$+$4054 (2.1\%) and J1826$+$1831 (8.8\%).  In both cases
the polarized emission is found in jet components on the stronger
side of the center of activity.

\end{abstract}


\section{Introduction}

Since being introduced 10 years ago as a subset of the ``compact
double'' class of objects \citep{phi80, phi82}, Compact Symmetric
Objects \citep[CSOs;][]{wil94} have yielded insights into the
evolution of active galactic nuclei (AGN) and unified models of AGN.
They have emission on both sides of the central engine, and their
small sizes ($<$~1~kpc) are thought to be due to their youth
\citep[$\leq$10$^4$~yr;][]{rea96a, ows98a, ows98b}.  CSOs may evolve
into Fanaroff-Riley II radio galaxies \citep{fan74, rea96b, pol03}, or
they may die out long before reaching that stage.  Alternatively,
\cite{car94, car98} have suggested that these objects are not small by virtue 
of youth, but rather are frustrated jets impeded by a dense interstellar 
medium.  With this model, they derive ages from 10$^6$ to 10$^7$ yr.  
The random
orientations of CSOs test the unified schemes of AGN that require an obscuring
torus of gas and dust surrounding the central engine.  The angle that
the line-of-sight of the observer makes with this torus determines
what features are seen \citep{ant93}.  CSOs often exhibit very broad
HI absorption lines \citep{tay99, pec00b}, and free-free
absorption \citep{pec99}, providing evidence for the existence of this
circumnuclear torus.  CSOs also typically exhibit very low
polarization, if any, due to the large Faraday rotation measures
induced by ionized gas associated with the torus.

The ages of these objects can be measured by hot spot advance
speeds.  That is, the speeds of the outward moving hot spots are
measured and the motions interpolated back to the center of the CSO,
assuming a constant source expansion rate.  These age estimates
are in agreement with, but more precise than, spectral 
aging arguments \citep{rea96a,mur99}.
Kinematic age estimates carried out by \cite{tay00}, \cite{ows98a},
and \cite{ows98b} are between 350 and 2000 years due to speeds of
approximately 0.1~c.  Jet components, however, exhibit relativistic
speeds \citep{tay00} as they feed the hot spots. 

In this paper, we attempt to classify the remaining CSO candidates in the COINS
sample \citep[CSOs Observed in the Northern Hemisphere;][]{pec00a} and
characterize the proper motions of hot spots and jet components of the
CSOs in the sample.  This allows us to increase the known CSO
population and to add to the growing body of CSO age estimates using this
kinematic method.  Of the 21 sources listed in Table 1, 10 have been
confirmed as CSOs by \cite{pec00a}.  The remaining 11 candidates have
been carefully examined, and 4 are rejected because they are
core-jet sources, while 5 are confirmed as CSOs based on the evidence
presented in this paper (J0000$+$4054, J0620$+$2102, J1111$+$1955,
J1143$+$1834, and J2203$+$1007), and 2 sources remain candidates 
pending further observations.  We also present the detection of
polarization in two CSOs, J0000$+$4054 and J1826$+$1831, a first for
the CSO class.

%Figures 1 and 2 display total intensity images for all of the sources
%discussed in this paper.  Figure 3 shows the polarized intensity for
%the two CSOs J0000$+$4054 and J1826$+$1831.  Figures 4 and 5 show
%changes in positions and fluxes, respectively, over time for
%J1415$+$1320.  A histogram of CSO ages known to date is presented in
%Figure 6.  Figure 7 shows the total flux of CSOs in this paper as they
%change with time.  Identifications, sky positions, and redshifts are
%given for each target source in Table 1.  In Table 2, properties of
%the CSO and candidate images are given, along with the current
%classification of each source.  In Table 3, we present estimates of
%the properties (flux densities, spectral index, proper motions, and
%polarized intensity) of dominant components within each source.  Table
%4 lists component and total fluxes for J1415$+$1320, which has two
%extra epochs of data from \cite{per96}.  Tables 5 and 6 provide
%further information on the proper motions and ages of components in
%the CSOs and candidates.  Tables 7 and 8 list the image parameters and
%component modelfits for the sources that are no longer candidates
%because they are core-jets.

In addition to their scientific appeal, CSOs can also be quite 
useful calibrators.  CSOs have been shown
to be remarkably stable in flux density \citep{fas01, and04}, making them
ideal sources to use as amplitude calibrators in monitoring
experiments, such as measuring the time delay between components of
gravitational lens systems.  CSOs to date have been observed to
have very low fractional polarization \citep{pec00a}, making them
useful to solve for leakage terms, or as an independent check
on the quality of the polarization calibration.

Throughout this discussion, we assume H$_{0}$=71 km s$^{-1}$
Mpc$^{-1}$, $\Omega_M$ = 0.27, and $\Omega_{\Lambda}$= 0.73.  Linear
sizes and velocities for sources with known redshifts have been
calculated using E.L. Wright's cosmology calculator
\footnote{http://www.astro.ucla.edu/$\sim$wright/CosmoCalc.html}.

\section{The Sample}

The CSOs and CSO candidates presented here are from the COINS sample
\citep{pec00a} and were selected as candidates from the VLBA
Calibrator Survey \citep[VCS;][]{bea02}. Identifications, sky
positions, and redshifts are given for each target source in Table 1.
Such large surveys are used for candidate selection because CSOs are rare
\citep[$\sim$5$\%$ of compact objects;][]{pec00a}.  It is also
necessary to go to moderate flux density levels ($\sim$100~mJy at
5~GHz) in order to form a complete sample.  The jet axis in each of
these objects usually lies within 60\deg\ of the plane of the sky, and
the hot spot motions are moderately slow (0.1 c) so relativistic
beaming does not play a major role.  CSOs are chiefly identified by
(1)~their high degree of symmetry, (2)~a compact, variable, flat
spectrum core in the center of the object, and (3)~proper motions that
are consistent with a centrally located core \citep{pec00a, tay00}.

\section{Observations and Analysis}

The observations were carried out at 8.4~GHz on 28 December 1997, 16
March 1998, 25 March 2000, and 2 December 2002, using the VLBA and one
VLA antenna\footnote {The National Radio Astronomy Observatory is
operated by Associated Universities, Inc., under cooperative agreement
with the National Science Foundation.}.  Observations were also made
of the 11 candidates at 15~GHz on 4 January 2001 using the VLBA.
Approximately half of the sources were observed in 1997 and
the other half in 1998, and so these are both referred to as the first
epoch at 8.4~GHz.  The results from this first epoch were presented by
\cite{pec00a} and are used here for the analysis of proper motions,
although the data from 1997 were recalibrated to achieve better
sensitivity.  Unavailable antennas in each epoch at 8.4~GHz were North
Liberty in 1997, Los Alamos in 1998, and St. Croix in 2002.  Amplitude
calibration was derived using measurements of the system temperatures
and antenna gains.  Fringe-fitting was performed with the AIPS task
FRING on the strong calibrator 3C~279.  Feed polarizations of the
antennas were determined at 8.4~GHz using the AIPS task LPCAL and the
unpolarized source J0427$+$4133.

Absolute electric vector position angle (EVPA) calibration was
determined by using the EVPA's of 3C~279
listed in the VLA Monitoring
Program\footnote{http://www.vla.nrao.edu/astro/calib/polar/}
\citep{tmy00} and checked with the polarization calibrator 1310$+$323.
No attempt has been
made to correct the electric vector polarization angles for
Faraday rotation, which is often significant on the parsec scale
for AGN \citep{zav03}.

For each source, the 15~GHz data was tapered to produce an image at
comparable resolution to the full resolution 8.4~GHz image.  The two
images were then combined to generate a spectral index map.  It is
important to note that spectral index maps made from two datasets
with substantially different ($u,v$) coverages may suffer from
significant systematic errors, especially in
regions of extended emission.

\section{Results}

Figure 1 displays total intensity images for all 17 of the CSO 
and CSO candidate sources discussed in this paper.  In all cases,
the 2000.227 epoch is used because it is the most sensitive.  See Table 2 
for a summary of the properties of
the CSO and candidate images, along with the current
classification of each source.
In the following section, we discuss results confirming 5 sources as
CSOs from our initial 11 candidates identified from the COINS
sample. 
Two sources are still regarded as CSO candidates.  Ideally, a
compact, flat spectrum, variable core component situated between two
steep spectrum components must be identified before sources can be
confirmed as CSOs.  For some CSOs with a jet axis very close to the
plane of the sky, the core may be so weak as to be undetectable, yet a
CSO identification can still be secured if there are symmetric,
edge-brightened, steep spectrum hot spots, extended lobes, or motions
consistent with a central core.  This was the case for all 5 of the
new CSOs.  On the basis of morphology and spectral index (defined
$S_{\nu} \propto \nu^\alpha$), we reject 4 candidates because they are
core-jets.  The maps of these sources are presented in Fig. 2
and their parameters are described in Table 3. 

For each CSO and candidate source, we provide in Table 4 the flux density of each component
at each of the three epochs.  We also provide an estimate of the 
core fraction (defined as the flux density measured in the core 
component divided by the total flux density measured in the VLBI 
image (see Table 2).  The spectral index between 8.4 and 15 GHz is also 
given, along with the linearly polarized flux observed in the most
sensitive second epoch.  

Relative proper motions (see Table 4) were difficult to detect in most of these
sources because of the slow apparent movement of CSO components.
However, where large errors are present, upper limits can be estimated for the motions,
and from those we can obtain interesting lower limits on the
kinematic ages.  Where redshifts are available, distances can be 
given in parsecs and velocities in terms of c.  Where identifiable, the core
is used as the reference for relative motions.  If there is no
identifiable core, the brightest hot spot is used as the reference
component.  Uncertainties in the positions were derived from the signal-to-noise
ratios for a given component and the synthesised beam. Motions are only given 
if the component motion is well
fitted by a weighted least squares fit where the slope of the line is
the velocity.  That is, a $\chi^{2}$ of 2 or below is acceptable for a
three epoch fit.  In cases where only one component of the velocity
was a good fit, either $x$ or $y$, only that velocity component is
used.  This is apparent from the position angle being either $-$90, 0,
or 90 degrees.  In most cases, this occurs when the source axis lies
close to the $x$ or $y$ axis.

Table 5 lists component and total fluxes for J1415$+$1320, which has two
extra epochs of data from \cite{per96}.  Tables 6 and 7 provide
summaries of the proper motions and ages of components in
the CSOs and candidates.  Finally, Table 8 lists the 
component modelfits for the core-jets sources.

\subsection{\bf CSOs and Candidates}

% !!!!!!!!! FOR REFS FOR Z, SEE WHICH REF W/I A REF IT COMES FROM! !!!!!!

\subsubsection{\bf J0000{\tt +}4054 (CSO)}  Also known as 4C~40.52,
this source has a central, compact group of components and extended
lobes to the north and south.  It has been identified with a galaxy of magnitude 21.4 \citep{sti96}.  It was given a tentative CSO
classification based on 1.6~GHz morphology by \cite{dal02}.  Although
no core could be determined from the spectral index map, the
morphology and proper motions of the components make its new CSO
classification quite clear.  The hot spot separation speed (motion of component
A with respect to component C) is $<$0.144 mas yr$^{-1}$ which gives
an age of $>$280 yr.  The core may be obscured by an optically thick
torus, leaving only the base of one of the jets visible (components B1
and B2).  This CSO also has detectable polarization in the southern
hot spot (component C) at 8.4~GHz (0.9 mJy, or 9$\sigma$; see Fig.~3a).  
This is considered the first confirmed CSO to show polarized
intensity.
%J0000+4054 CSO some pol, no core in spec indx? motion, vary consistent CSO
% 4C 40.52  G (mag 21.4 stickel & Kuer 96) 

\subsubsection{\bf J0003{\tt +}4807 (CSO)}  This CSO has a double-lobe
morphology with a faint central component.  This component, labeled B,
is thought to be the core \citep{pec00a}.  The motion of hot spot A
away from the core provides an upper limit
for the hot spot advance speed ($<$0.014 mas yr$^{-1}$),
which gives a lower limit for the age of 340 yr.  
%J0003+4807 COINS CSO

\subsubsection{\bf J0132{\tt +}5620 (Candidate)}  This AGN is still
considered a candidate CSO because of its double-lobe structure.
However, the core is not clearly identifiable either from the spectral index
map or component variability (see Table 4).  Component D1 appears
to move with marginal significance to the west with respect 
to D2, but otherwise no motions are detected.    

%J0132+5620 Core-jet? spec index-core at right, but CSO motion, vary? 

\subsubsection{\bf J0204{\tt +}0903 (CSO)}  With a bright central
core, this CSO has a jet extending in the eastern direction and a
fainter counter-jet southwest of the core.  It initially seems as
though components A and B are moving slowly towards the core
(component C).  However, it is possible that the core is
ejecting components to the east and west, but the new eastern
component is Doppler boosted and makes the core appear to shift
in the direction of the stronger jet.  The
motion of component D away from the core is 0.070$\pm$0.011 mas yr$^{-1}$, giving an
age of 240$\pm$36 yr.

%J0204+0903 COINS CSO; motion; mag >20 (Snellen et al. MNRAS 329
%700 2002). 

\subsubsection{\bf J0427{\tt +}4133 (CSO)}  A very bright and compact
object, this CSO was used to solve for the polarization leakage
terms.
The prominent and variable ($\sim$14\%) core is easily identifiable as
component B.  Component C displays a surprisingly large motion with respect to B in the
expected direction.  With a motion of 0.060$\pm$0.013 mas yr$^{-1}$, the
estimated age
of this component is only 20$\pm$4 yr.  If this is a hot spot, then this
makes this the youngest CSO known.  Alternatively, we could be measuring
a fast jet component in which case the absence of detectable hot spots
farther out is puzzling.

%J0427+4133 COINS CSO

\subsubsection{\bf J0620{\tt +}2102 (CSO)}  This nearly equal double source
has been identified here as a CSO by the steep spectral indices of its
hot spots ($\alpha \simeq -$1.0) and their lack of a significant hotspot separation velocity.
This allows us to put an upper limit on the hot spot advance speed of
$<$0.013 mas yr $^{-1}$, and a lower
limit on the age of 2060 yr.
%J0620+2102 CSO - spec indx, morph, motion, and vary   

\subsubsection{\bf J0754{\tt +}5324 (CSO)} This CSO is dominated
by two hot spots with
a strong jet leading towards the northernmost lobe.  With component C as the
reference, we find an upper limit on the hot spot advance
speeds of $<$0.009 mas yr$^{-1}$ for component A, providing a lower limit to
the age of 2220
yr.  The core identified by \cite{pec00a} (B) is not
consistent with the motions presented here.  Rather, this component appears to
be a jet component speeding towards the northwestern hot spot (C) at
0.060$\pm$0.026 mas yr$^{-1}$.  Component B most likely represents the
base of the jet with an obscured core to the immediate southeast.
%J0754+5324 COINS CSO; motion, mag >20 (Snellen)

\subsubsection{\bf J1111{\tt +}1955 (CSO)}  Identified with a galaxy of
magnitude 18.5 and a redshift of 0.299 \citep{pec00b}, we can now
classify this source as a CSO due to its edge-brightened lobes having steep
spectra ($\alpha \simeq -$0.9 and $-$0.8) and a separation speed
$<$0.013 mas yr$^{-1}$, or $<$0.19~c, where A is the
reference.  This gives an age limit of 1360 yr.    
%J1111+1955 CSO - spec, motion; mag 18.14 (Snellen); G mag 18.5
%z=0.299 (Peck 00b). 

\subsubsection{\bf J1143{\tt +}1834 (CSO)}    This source
is classified here as a CSO due to its steep spectrum double hot spots
  ($\alpha \simeq -$0.8).  With component B as the reference, the hot spots are
moving away from each other at a speed $<$0.010 mas yr$^{-1}$ for an age limit
of 690 yr.
%J1143+1834 CSO - spec, morph and vary, motion; mag >20 (Snellen)

\subsubsection{\bf J1311{\tt +}1658 (Candidate)} As a CSO candidate
  this object appears to have a central core (B1 or B2) with a jet extending to
  the northwest (C and D) and a faint counter-jet or hot spot to the
  southeast (A).  However, the spectral indices of all the components
  are flat ($\alpha \simeq -$0.2 and $-$0.4) and no reliable motions
are detected.  It is worth pointing out that this is a difficult source to
  modelfit because of its extended structure. 
%J1311+1658 ???;  mag >20 (Snellen)

\subsubsection{\bf J1414{\tt +}4554 (CSO)}  This CSO has been
associated with a galaxy of magnitude 19.9 and a redshift of 0.190 by
\cite{fal98}.  It has a bent northern lobe (B) and an edge-brightened
southern lobe (A).  There are no detectable motions within the upper limit of
0.014 mas yr$^{-1}$, or 0.14~c.  This gives a lower limit to the age
of 2030 yr.
%J1414+4554 COINS CSO; motions; G z=0.190 mag 19.9 (Peck00b); mag 18.6 z=0.1860 (Snellen)

\subsubsection{\bf J1415{\tt +}1320 (CSO)}  Also known as PKS~1413$+$135,
this CSO has also been classified as a BL Lac object due to its
optical properties \citep{bei81, bre81}.  This proved to be an unusual
object from the start as it was later found to be the first BL Lac to
be associated with a disk galaxy ($M_{v}=$ 19.6, $z=$ 0.25) from
optical images, HI absorption, and extinction in the near infrared
\citep{mch91, car92}.  This disk also gives rise to evidence of
molecular absorption in the line-of-sight of the nucleus \citep{wik94,
con99}.  However, \cite{sto92} suggested that the BL Lac could be a
background quasar that has been gravitationally lensed by the
foreground spiral based on the fact that there was no thermal IR
emission that would normally be expected for a spiral host.
However, much of the evidence to date confirms that the radio object
is a part of the optical galaxy, not lensed by it \citep{mch94, lam99,
per02}.

\cite{per94} imaged a core, jet, and counter-jet on VLBI scales at 1.6
and 8.4~GHz, putting the BL Lac classification in jeopardy with
regard to the unified schemes of AGN.  Later, \cite{per96} confirmed
at 1.4, 2.3, 4.8, and 8.4~GHz that no double images existed as
evidence of lensing and that the source most likely resembled the
members of the CSO class.  The northeastern counterjet vanishes
at high frequencies.

In addition to the epochs listed above, we used calibrated data from
\cite{per96} from July 1994 and 1995 for a longer timeline.  However,
our 2002.919 epoch could not be properly modelfit on this complex source
and was not used for the motion analysis.  The total flux was
drastically lower in 2000.227 than in the other 4 epochs ($\leq$60\%).
Most of this drop was in component C (see Fig.~4), confirming the
classification of this as the core with its extreme variability.
There was also a drop in the flux of the newest jet component, C3.
The core was used as the reference component for the proper motion
analysis.  The counter-jet component east of the core (B2) has a motion
over the 4 epochs at 0.060$\pm$0.024 mas yr$^{-1}$, corresponding to a
velocity of 0.80$\pm$0.30~c.  This gives an age of 130$\pm$47 yr (see
Fig.~5) for this component.  The jet component closest to the core
(C3), appears to be only 22$\pm$2 yr old due to a larger proper motion of
0.087$\pm$0.009 mas yr${-1}$, or a velocity of 1.10$\pm$0.11~c.  This
velocity, however, is only 25\% of that measured by \cite{kel98} at 15~GHz.

For two-sided sources it is often possible to obtain information about
the true jet velocities, including both speed and orientation (e.g.,
\nocite{tay97} Taylor \& Vermeulen 1997).  For simultaneously ejected components moving in
opposite directions at an angle $\theta$ to the line of sight at a
velocity $\beta$, it follows directly from the light travel time
difference that the ratio of apparent projected distances from the
origin ($d_{\rm a}$ for the approaching side, $d_{\rm r}$ for the
receding side) as well as the ratio of apparent motions (approaching:
$\mu_{\rm a}$, receding: $\mu_{\rm r}$) is given at any time by
\begin{equation}
{{\mu_{\rm a}}\over{\mu_{\rm r}}} = {{d_{\rm a}}\over{d_{\rm r}}} = 
\Biggr({{1+\beta\cos \theta}\over{1-\beta\cos \theta}}\Biggl)\,.
\end{equation}
Given the discrepant ages for C3 and B2, we cannot make the assumption
that they were ejected at the same time. If, however, we assume that 
their velocities are characteristic of jet and counterjet velocities
in PKS~1413$+$135, then we obtain $R$=1.45, or $\beta\cos \theta = 0.18$.
This requires that $\beta$ is greater than 0.18, and that $\theta$ is 
less than 80\deg.  The rapid, and extreme variability for the core 
component suggests a small angle to the line-of-sight, which could
be consistent with the Doppler boosting model if the jets are 
initially slow.   Why the jet should fade so quickly on the jet
side compared to the counter-jet side (Fig.~4) is unclear, and could
indicate that explaining asymmetries in terms of a simple Doppler 
model is invalid for this source.

%J1415+1320 COINS CSO; motion; PKS1413+135 QSO mag 20 z=0.25 (Peck00b); mag
%19.88 z=0.2467 (Snellen); BL Lac z=0.247 (Zensus et al. 02);
%radio-loud spiral? (Perlman et al. 02); ``disk dominated'' host
%galaxy - not the first for FR I (Lamer et al. 1999); disk with HI and
%moleuclar absorption in line-of-sight of nucleus (Conway 99); Perlman
%et al. 96 - 8GHz maps from Jul94, not rel beamed

\subsubsection{\bf J1546{\tt +}0026 (CSO)}  This bright, compact CSO
is associated with a magnitude 20 galaxy of redshift 0.55
\citep{hec94}.  It was classified as a complex morphology
compact steep spectrum (CSS) source by \cite{sta99} based on
their 4.8~GHz VLBI observations.
There is most likely missing extended emission on VLBI scales because
the total flux at 8.4~GHz for these three observations is lower than
the flux measured with the VLA at the same frequency \citep[663, 543, and 602
mJy with VLBA, this paper; 909 mJy with VLA,][]{bro98}.  The motion of
the westernmost component (C) is surprisingly towards the
central core (B) and is 0.054$\pm$0.008 mas yr$^{-1}$,
corresponding to a velocity of 1.10$\pm$0.17~c.  If we assume that
C is actually stationary or nearly stationary with respect to the
core, the motion of component B may be due to a new jet
component emerging at relativistic speeds and moving the position of
model component B towards C.  We also note that it is difficult to 
describe this very extended source with Gaussian components.  No
kinematic age estimate has been derived.
%J1546+0026 CSO spec  mag 18.45 z=0.55 (Snellen); G mag 20 z=0.55 (Peck00b)

\subsubsection{\bf J1734{\tt +}0926 (CSO)}  Another edge-brightened,
double hot spot CSO, this source is associated with a galaxy of magnitude 20.7
\citep{pec00b} and redshift 0.61 \citep{peru02}.  It was
tentatively classified as a CSO by \cite{sta99} based on its 4.8~GHz
morphology.  The hot spot advance
speed of component A from B is $<$0.008 mas yr$^{-1}$, or
$<$0.18~c.  This gives
a lower limit to the age of 1780 yr.
%J1734+0926 CSO spec. motion; G mag 20.7 (Peck00b); Perucho & Marti
%2002 z=0.61. 

\subsubsection{\bf J1816{\tt +}3457 (CSO)}  This north-south,
double-lobe source has been identified in the optical with a galaxy of
magnitude 18.7 and redshift 0.245 \citep{pec00b}.  Hot spot B is moving
with respect to A with a speed of 0.036$\pm$0.009 mas yr$^{-1}$, or
0.40$\pm$0.11~c in a direction almost perpendicular to the source
axis.  This side-to-side motion of the hot spots is not often seen in
CSOs \citep{pol03} although 1031$+$567 also shows this behavior
\citep{tay00}.  This may be due to an interaction with the ISM, as
cited for the case of 4C~31.04 by \cite{gir03}.  No age has been
determined for this source.  
%J1816+3457 COINS CSO; motion; G mag 18.7 z=0.245 (Peck00b)

\subsubsection{\bf J1826{\tt +}1831 (CSO)}  This large, extended CSO
has a bright hot spot and corresponding jet (C and D) to the west of
the core (B) and a fainter hot spot to the east (A).  Although A is
only 16\% as bright as D, this is within the 10:1 brightness
ratio that is accepted as the maximum for CSO classification
\citep{tay03}.  This CSO displays the most significant polarization of
any CSO to date with 3.0 mJy in component C ($\sim$8.8\%; see Fig.~3b).  This component is also moving away from the core at a speed of
0.037$\pm$0.009 mas yr$^{-1}$, giving it a component age of 380$\pm$93 yr.
Component D is moving at a speed of 0.013$\pm$0.006 mas yr$^{-1}$ and
gives a source age estimate of 3000$\pm$1490 yr.
%J1826+1831 COINS CSO with pol! motion

\subsubsection{\bf J2203{\tt +}1007 (CSO)} This new member of the CSO
class has two lobes with
an extension (B) on the eastern lobe (A) pointing towards the geometric
center of the source.  All of these components have steep spectral
indices (A and B, $\alpha \simeq -$0.8; C, $\alpha \simeq -$1.3).  The
overall expansion of the source is measured for component C with respect to A1
with a speed of $<$0.011 mas yr$^{-1}$, or an age of $>$940 yr.
Although the 22.53
magnitude optical component is  classified as a QSO, it has recently
been shown to be extended in the optical \citep{dal02b}.
%J2203+1007 CSO motion, vary;  mag >20 (Snellen); mag 22.53 and optically
%extended (Dallacasa et al. 2002)

\subsection{\bf Rejected Sources}

\subsubsection{\bf J0332{\tt +}6753} This large, extended source has a
flat spectrum, compact core to the north (component C; $\alpha \simeq
-$ 0.1) with three steep spectrum, extended components approximately
40 mas to the south (B, A, and an unmodeled component; $\alpha \simeq
-$0.8 and $-$0.5, respectively).  Component A has 1.7 mJy of polarized
flux, which is typical for a jet component.  Components A and B are
moving away from C at less than 0.022 and 0.014 mas yr$^{-1}$,
respectively.
%J0332+6753 core-jet - based on spec indx, compact, motions, pol

\subsubsection{\bf J0518{\tt +}4730} We identify component A, the easternmost component, as the core due to its compact
morphology and large variability.  It is the reference for the
motions of the steep spectrum jet components labeled B and C.
These were measured to be moving at 0.026$\pm$0.009 west and
0.018$\pm$0.009 mas yr$^{-1}$ southwest, respectively.  Component C is polarized
with 0.8 mJy of flux.
%J0518+4730 core-jet - motion, pol, morph and vary, but not spec indx

\subsubsection{\bf J1311{\tt +}1417} This source is an interesting
bent core-jet source, but is no longer a CSO candidate because its
northernmost component (C) has been identified as the core with a
spectral index of $\alpha \simeq -$0.3.  Jet components A and B have
steep spectral indices ($\alpha
\simeq -$1.0) and polarized flux (4.4 and 9.0 mJy, respectively).
Their motions with respect to C, however, were plagued with large
errors and could not be determined.  This source has been 
associated with a QSO of magnitude 19.5 and redshift 1.952 by
\cite{pec00b}.
%J1311+1417 bent core-jet - spec, vary, pol; QSO mag 19.5 z=1.952 (Peck00b)

\subsubsection{\bf J2245{\tt +}0324} Component B is the core of this
source with a spectral index of $\alpha \simeq -$0.2.  Although it has
a very faint counterjet (too faint to be modelfit), it does not
contain enough flux as compared to the jet (A) in order to classify it
as a CSO (brightness ratio $\sim$20:1).  Component A is slightly
steeper than the core with $\alpha \simeq -$0.5 and is moving away
from B at a speed of 0.044$\pm$0.007 mas yr$^{-1}$, or
1.20$\pm$0.19~c.  This velocity is also more consistent with a
core-jet than a CSO.  Like J1546$+$0026, this source is missing flux
on VLBI scales compared to VLA observations at the same frequency by
\cite{bro98}.  Our observations have total fluxes of 389, 368, and 368
mJy, whereas the VLA observations have a total flux of 624 mJy.  This
radio source is associated with a QSO of magnitude 19 and redshift
1.34 \citep{wol97}.
%J2245+0324 core-jet spec, pol, motions, vary; QSO mag 19 z=1.34
%(Wolter et al. 1997); missing alot of flux compared to VLA

\section{Discussion}

The kinematic ages for the sources measured in this sample range from 20$\pm$4 to
3000$\pm$1490 yr, a slightly larger range than that found in the literature 
(see $\S$1).  If the definite age estimates in this sample are combined
with the ages of 10 sources cited by \cite{pol03}, 7 out of the total
13 sources are under
500 yr old, and when lower limits are included, 9 of 23 lie in this range
(see Fig.~6).  This is surprising since our naive expectation for
a steady-state population would be a uniform distribution of ages.
It may be that CSOs undergo phases of activity that only
last for several hundreds of years.  Low frequency observations could 
reveal remnants of earlier stages of activity, as in the 
case of the compact double 0108$+$388 \citep{bau90}.  There also may be a selection effect
working against older CSOs.  Their apparent motions may be so slow as
to be undetectable within reasonable limits on the time baselines that
are used.  Although it has been shown that AGN grow dimmer with size on large scales, these sources are 
thought to increase in luminosity up to a size of 1 kpc \citep{sne00, ale02}.  
Therefore, a luminosity-based selection effect can probably be ruled out.  There is
also a theory that many CSOs ``fizzle out'' after a short period of
activity \citep[and references therein]{pol03}.  This age distribution may
be evidence of such a process.
However, larger samples and better estimates are needed to determine
the true age distribution.

An extreme example of this apparent youth is J0427$+$4133.  By its
age estimate (20$\pm$4 yr), it may indeed be a very young object, in which 
case this would seriously constrain models of jet formation.  It may also be
evidence for a new phase of activity.  The fast moving component 
in that interpretation is then 
more likely to be a part of the jet and not a hot spot.  Another possibility
is that the way for the jet has been cleared and there is no material
against which it could form a hot spot.  The source has only
been detected as a point source at 1.4~GHz with the VLA \citep[The
NRAO VLA Sky Survey;][]{con98}.
Low frequency VLBI observations might still 
be able to confirm the presence of more distant extended structure, 
on sub-arcsecond scales.

%$\bullet$ Look at spread in CSO ages (incl Polatidis and Conway '03)

%$\bullet$ Smaller ones only show jet, not hot spots? (J0427 and J1546)

In the past, CSOs have been observed to be unpolarized.  However, with
the high dynamic range achieved in epoch 2000.227, polarized flux was 
detected in J0000$+$4054 (2.1\%) and J1826$+$1831
(8.8\%).  The polarized components in each of these sources appear to
be on the side with the more prominent jet.  In this orientation, the polarized components
might be free of obscuration by the central torus.  Under this model,
it is unlikely that polarization would be seen on both sides of a CSO
core.  It would be interesting to determine the Faraday rotation
measures for the first time in CSOs.  For completeness we note that
two CSO candidates in the southern sky also show polarization
(J0204$-$2132 and J1419$-$1928) and one CSO shows a possible detection
of polarization (J0735$-$1735, 0.2\%) in \cite{pec02}.

%$\bullet$ Why detect polarization? viewing angle and distance from
%supposed torus

For the 6 CSOs with identifiable cores, the percentage of the total
flux at 8.4~GHz that is attributed to the core ranges from 4 to 86\% (See Table
4).  In three of these (J0204$+$0903, J1415$+$1320, and J1826$+$1831),
there is is a clear difference in the length and brightness of the jet
and counter-jet of each, suggesting an orientation close to the 
line-of-sight.   Two of these sources (J0427$+$4133
and J1546$+$0026) are compact, and the final one (J0003$+$4807) has
only a weakly visible core (4\% the total flux).  This suggests that
certain conditions must exist for the core to be visible, such as jet
axis orientation.

%$\bullet$ Core prominence

On VLA scales, variability of CSOs has been shown to be constant to
within 1\% over time periods of 1 week to 10 months \citep{fas01}.
Although our experiment was not set up to accurately monitor the
total fluxes, the variation stays under 10\% for most of the sources
in the sample
(Fig.~7), and is likely to be instrumental in origin.  Some of the
outliers include the core dominated source J1415$+$1320 (see Fig.~4) and
J1546$+$0026, the CSO candidates J1311$+$1658 and J0132$+$5620, and
the CSOs J1816$+$3457, J1826$+$1831, and J2203$+$1007.  It is expected
that core-dominated CSOs would vary, and the variability of the two
CSO candidates could indicate that they are likely to be either
core-dominated CSOs, or core-jets.  It is not understood why three CSOs
in this sample vary by as much as 30\% over a five year time
period, though no careful studies of CSO variability have yet
been carried out to explore this time-scale. 

%$\bullet$ More data for z, motions

\section{Conclusions}

We have detected motions in the hot spots of 3 sources, yielding new
kinematic age estimates.  We also find motions in both the jet and
counterjet of the unusual CSO J1415+1320 (PKS 1413+135).  We place
limits on the kinematic ages for 10 more sources.  When combined with
kinematic age estimates from the literature we find that 7 of 13
sources are under 500 years old.  Longer time baselines are needed to
detect motions of the slow moving components, primarily the hot spots,
in CSOs and to accurately determine the ages of the objects for which
only limits have been determined.  For these CSOs, ranging in redshift
from 0.19 to 0.61, the sampling interval chosen of 2 years was a bit
short, with most sources appearing very nearly identical from epoch to
epoch. It is worth noting that the apparent expansion velocity of the CSO
PKS 1934-638 over a time baseline of 32.1 yr is similar. \citep{ojh04}.  
The completeness of the identification of redshifts for the
sample ($\sim$ 50\%) could also be improved upon and would allow for
the determination of projected velocities and linear sizes.

In the course of our study we discovered linear polarization in the
jets of the CSOs J0000$+$4054 (2.1\%) and J1826$+$1831 (8.8\%). These
are the first confirmed CSOs to be detected in polarization.  In both
cases the polarization is found on the side of the stronger jet.  This
is consistent with unified schemes since the Faraday depth on the
approaching side should be less, though still significant.  Sensitive,
multi-frequency VLBI observations of these sources to determine their
Faraday rotation measures should be carried out.

\acknowledgments 

We thank Eric Perlman for providing calibrated 8.4 GHz VLBA
observations from July 1994 and 1995 for J1415+1320.  NEG gratefully
acknowledges support from the NSF REU program. ABP and MG thank the
NRAO staff for their hospitality during part of this research.  We
also thank an anonymous referee for constructive comments.  This
research has made use of the NASA/IPAC Extragalactic Database (NED)
which is operated by the Jet Propulsion Laboratory, Caltech, under
contract with NASA.

\clearpage

%Figure 1 -- CSOs and candidates
\begin{figure}
\figurenum{1}
%\vspace{20.4cm}
%\special{psfile=f1a.ps hoffset=-30 voffset=0 hscale=80.0 vscale=80.0}
\epsscale{0.84}
\plotone{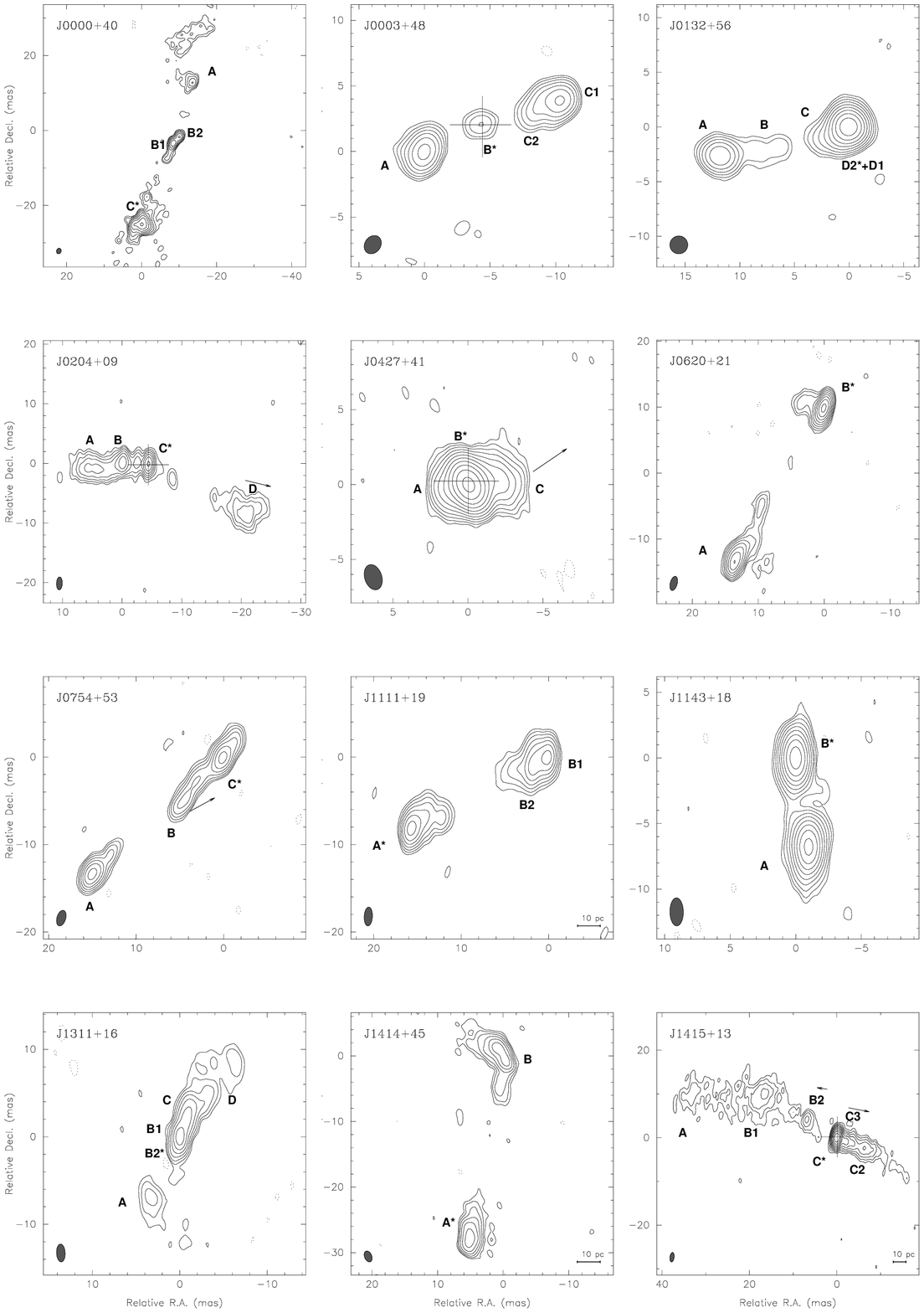}
\caption{ Total intensity contours of the CSOs and remaining candidates
  in the COINS sample at 8.4~GHz.  Where redshifts are available, a
  10~pc scale is indicated.  A cross indicates the location of the
  core, if identifiable.  Component labeling is consistent with Peck
  \& Taylor 2000.  Arrows indicate the direction of motion where
  applicable and are magnified 5 times for clarity.  An asterisk(*)
  denotes the reference components.  The synthesized beam is in the bottom-left corner
  of each image.  Image parameters are given in Table~2.}
\end{figure}
\clearpage

%Figure 1 -- CSOs and candidates STILL
\begin{figure}
\figurenum{1}
%\vspace{20.4cm}
%\special{psfile=fig1b.eps hoffset=-30 voffset=0 hscale=80.0
%  vscale=80.0}
\epsscale{0.95}
\plotone{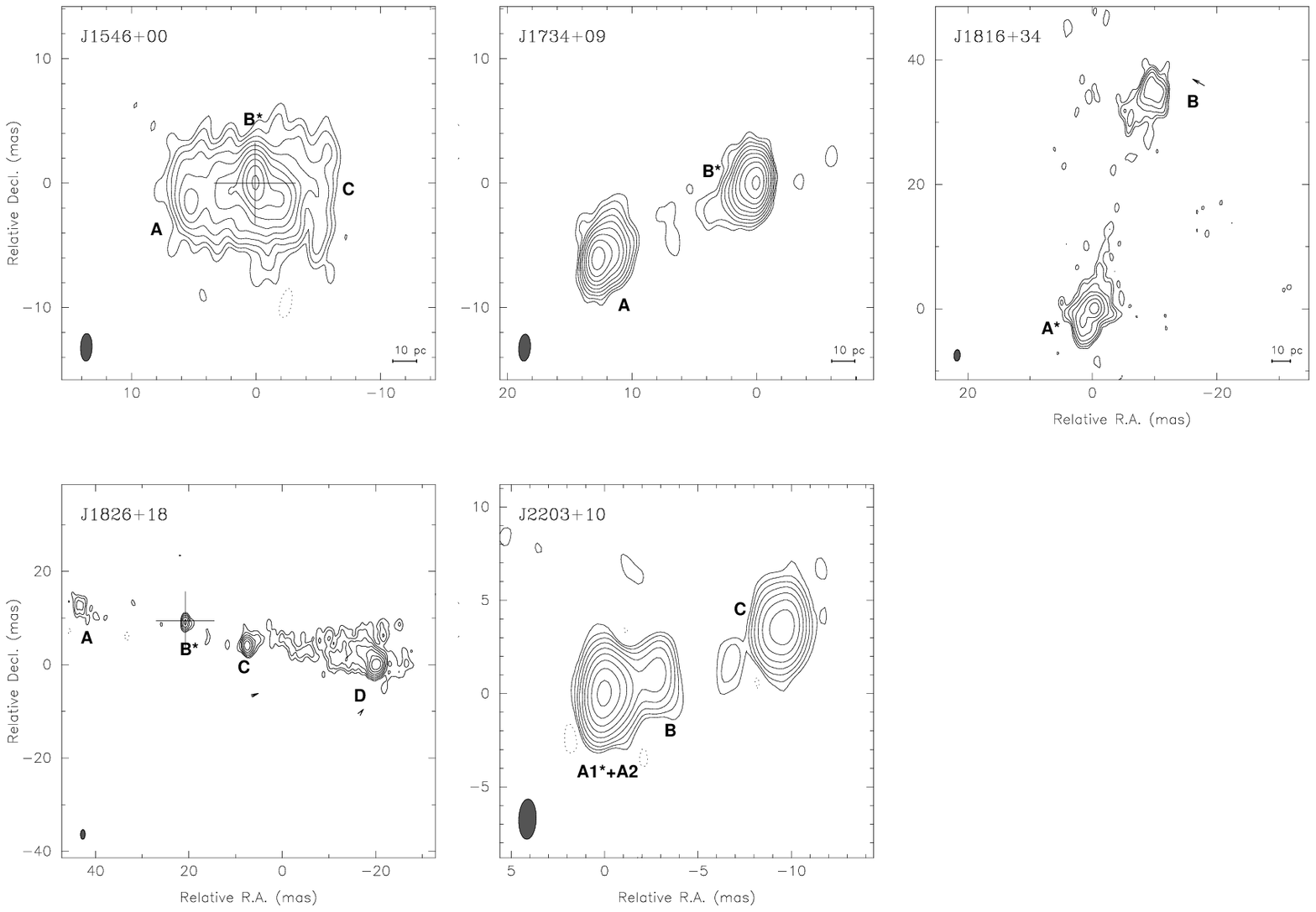}
\caption{ Continued.}
\end{figure}
\clearpage

%Figure 2 -- Core-jets
\begin{figure}
\figurenum{2}
%\vspace{20.4cm}
%\special{psfile=f2.eps hoffset=-30 voffset=0 hscale=80.0
%  vscale=80.0}
\epsscale{0.95}
\plotone{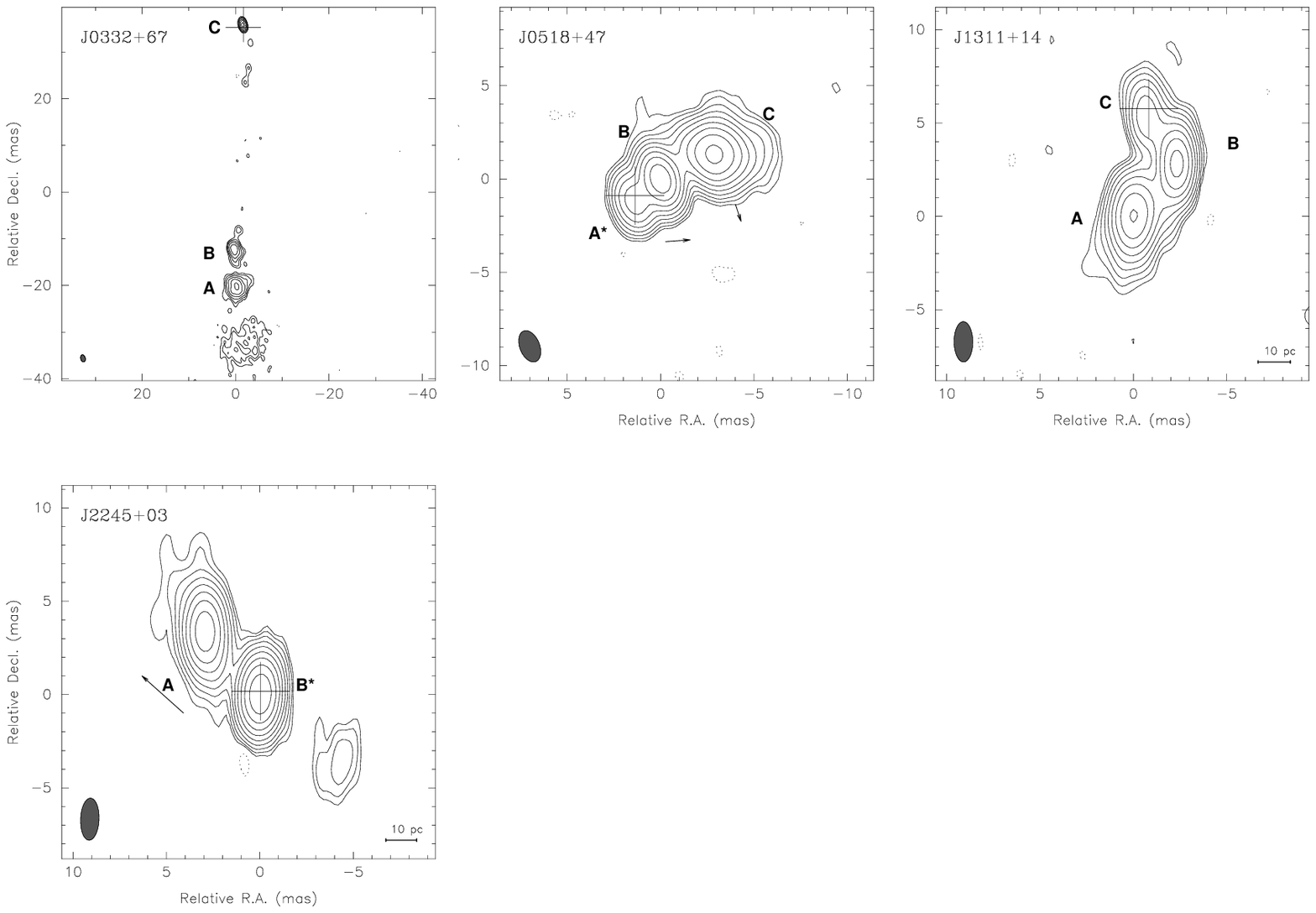}
\caption{ Total intensity contours of the sources rejected from the COINS sample at 8.4~GHz.  These are now classified as core-jets rather than CSOs.  Where redshifts are available, a
  10~pc scale is indicated.  A cross indicates the location of the
  core.  Component labeling is consistent with Peck
  \& Taylor 2000.  Arrows indicate the direction of motion where
  applicable and are magnified 5 times for clarity.  An asterisk(*)
  denotes the reference components.  The synthesized beam is in the bottom-left corner
  of each image.  Image parameters are given in Table~3.}
\end{figure}
\clearpage

%Figure 3 -- CSO polarization
\begin{figure}
\figurenum{3}
\vspace{20.4cm}
%\special{psfile=J0000.pol.ps hoffset=0 voffset=140 hscale=60.0
\includegraphics{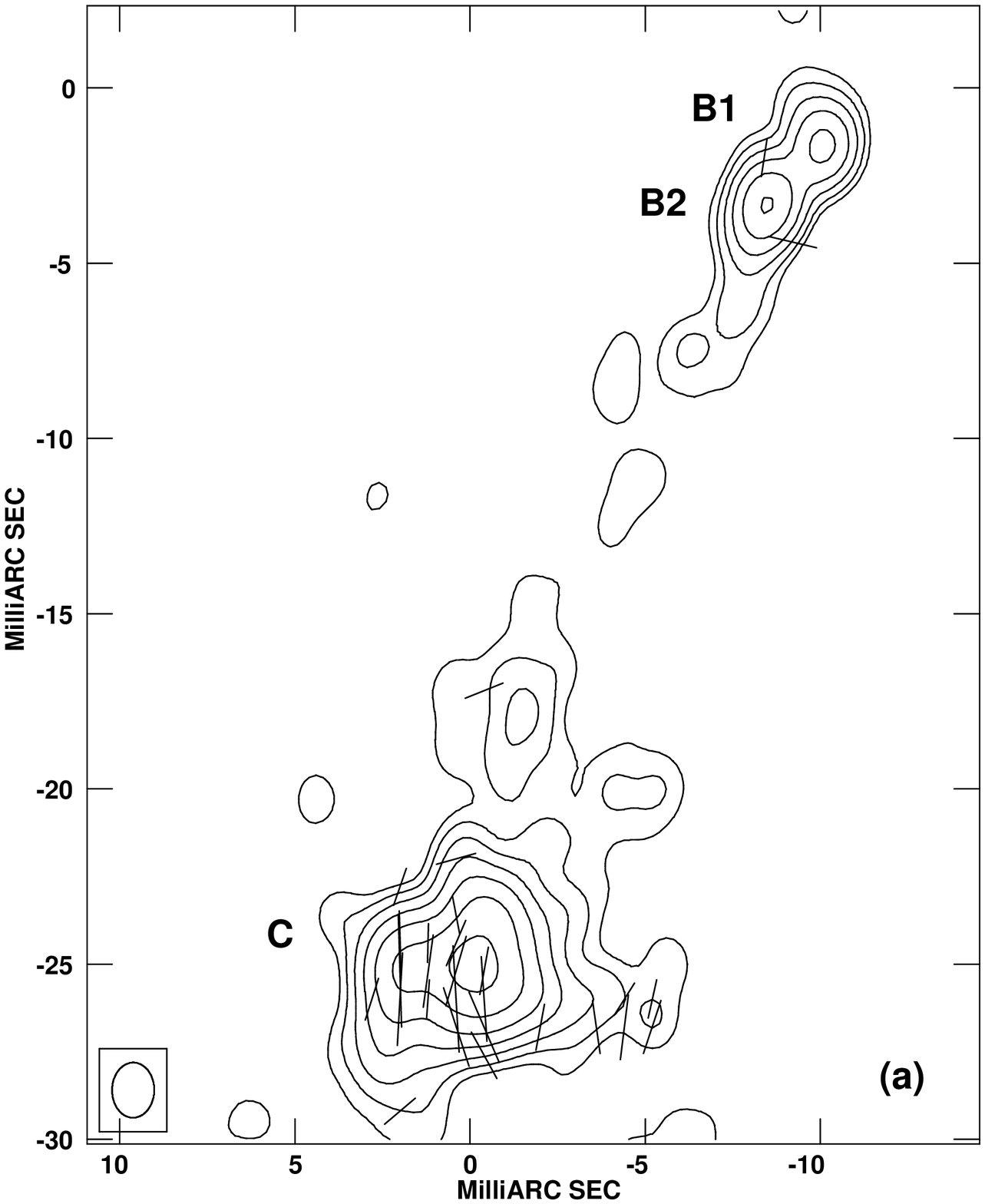}
%\special{psfile=J1826.pol.ps hoffset=0  voffset=-140 hscale=60.0
\includegraphics{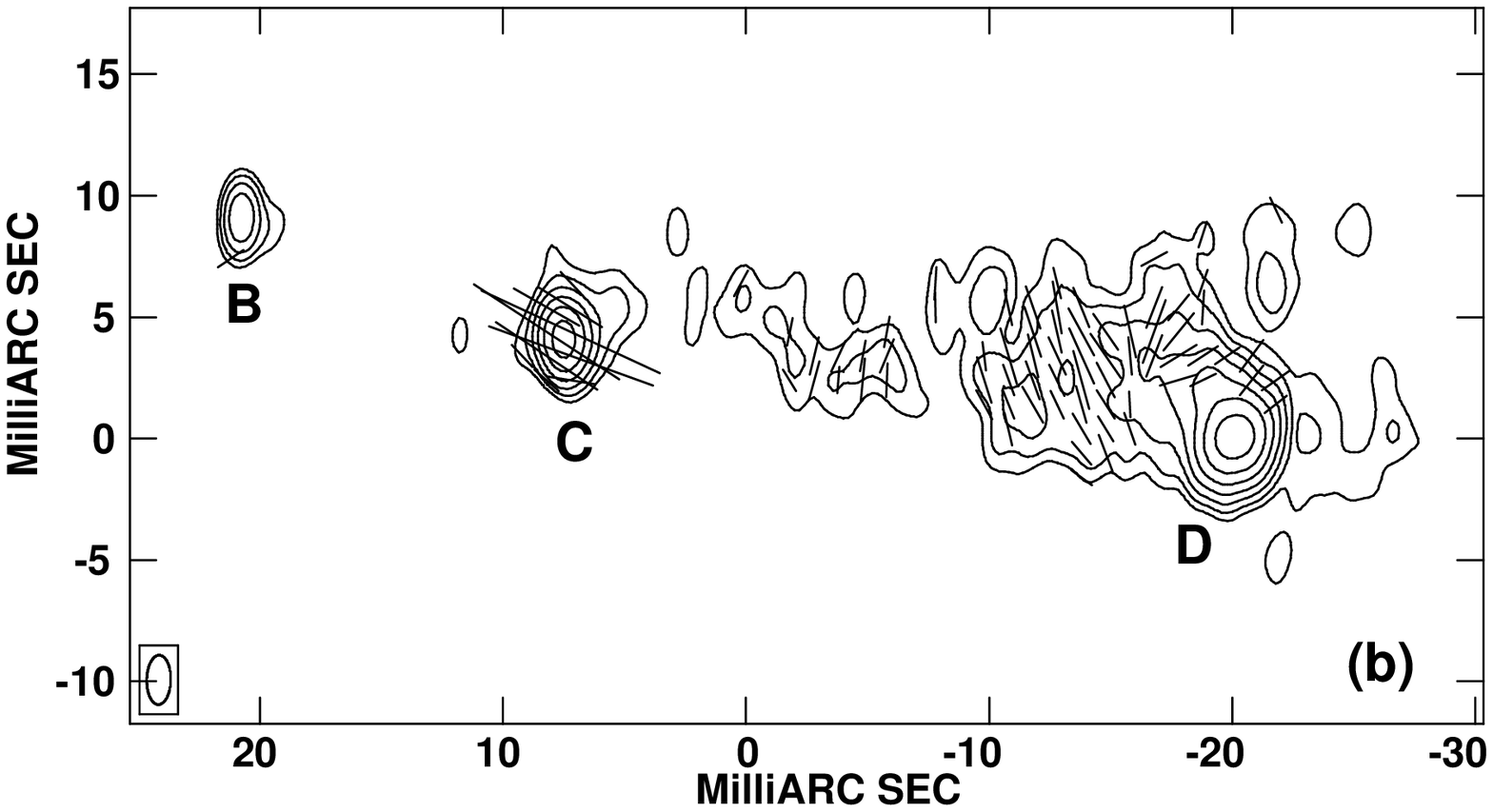}
\caption{Total intensity contours of {\bf (a)}J0000$+$4054 and {\bf (b)}J1826$+$1831 at
  8.4~GHz with electric polarization vectors superimposed.  Component A has been omitted in each plot as the polarized components have been enlarged.  A vector
  length of 1 mas corresponds to a polarized flux density of 0.25 mJy/beam.
  Contour levels begin at 0.5 mJy/beam and increase by factors of 2 for
  both sources.}  
\end{figure}
\clearpage

%Figure 4 -- J1415 Component fluxes
\begin{figure}
\figurenum{4}
\vspace{20.4cm}
\includegraphics{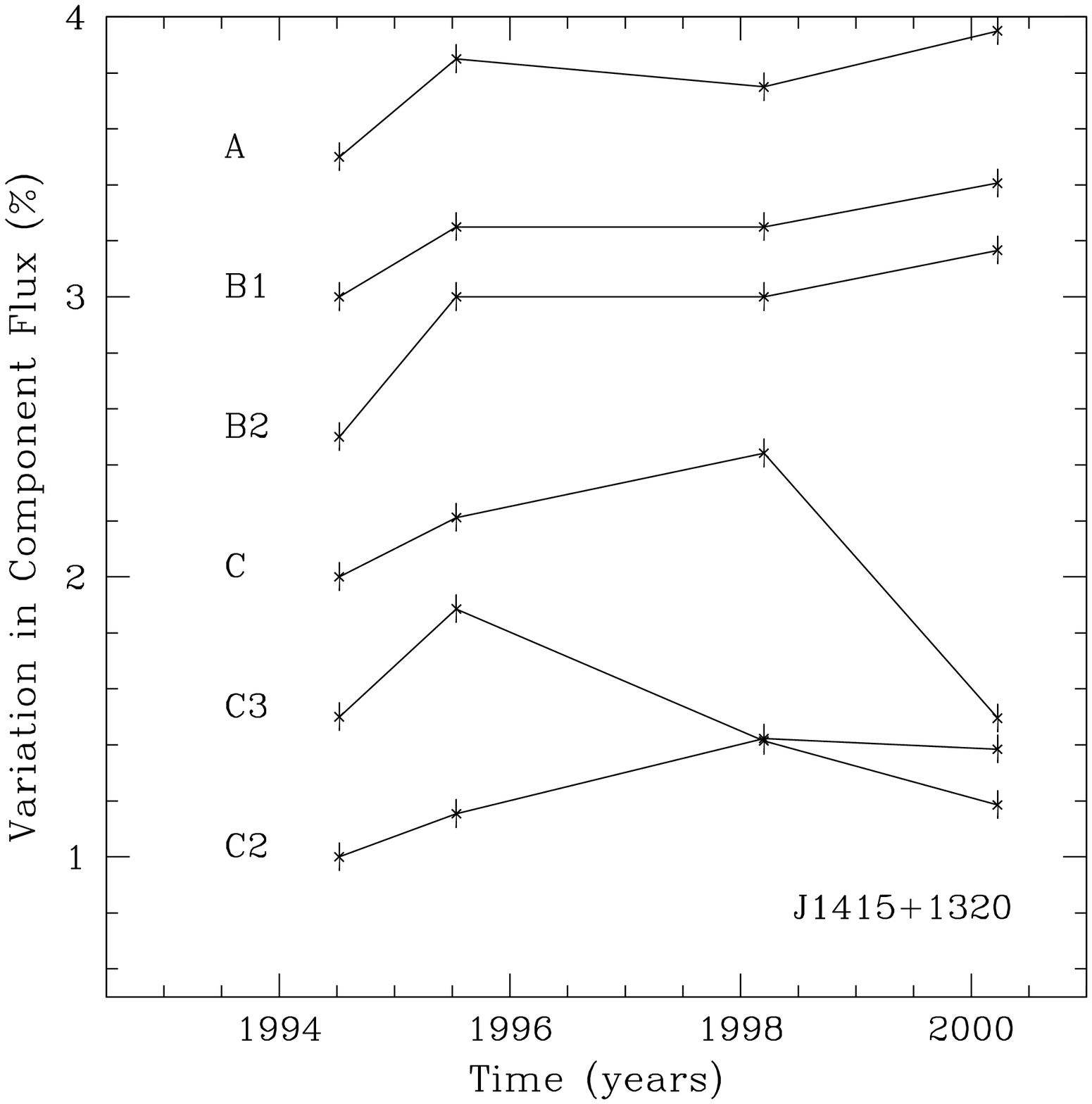}
\caption{Changes in the component fluxe densities of J1415$+$1320 over the four
  epochs used in the proper motion study.  After C2, each component
has been offset by 50\% to improve legibility. Note the large drop in flux
  of the core and innermost jet component.}
\end{figure}
\clearpage

%Figure 5 -- J1415 Overlay
\begin{figure}
\figurenum{5}
\vspace{20.4cm}
\includegraphics{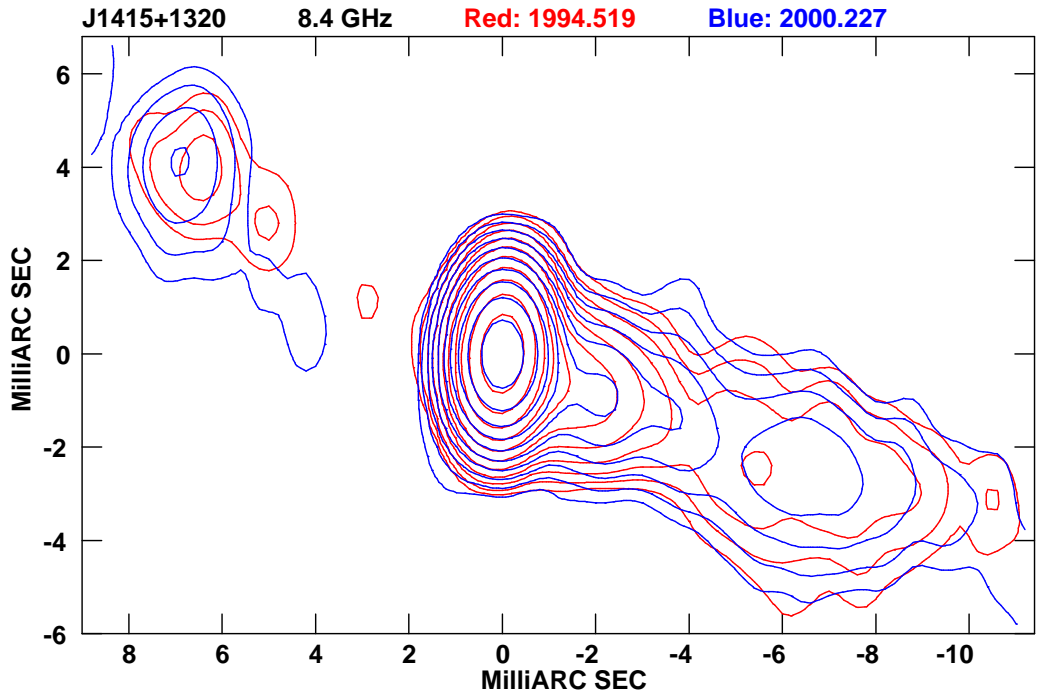}
\caption{Overlay of total intensity contours of J1415$+$1320 in
  1994.519 (red contours) and
  2000.227 (blue contours).  The motion of B2 (upper left corner) is
  apparent.  However, the confusion of components in the jet can also be
  seen.  This made modelfitting and obtaining proper motion measurements difficult.}
\end{figure}
\clearpage

%Figure 6 -- Histogram of Ages
\begin{figure}
\figurenum{6}
\vspace{20.4cm}
\includegraphics{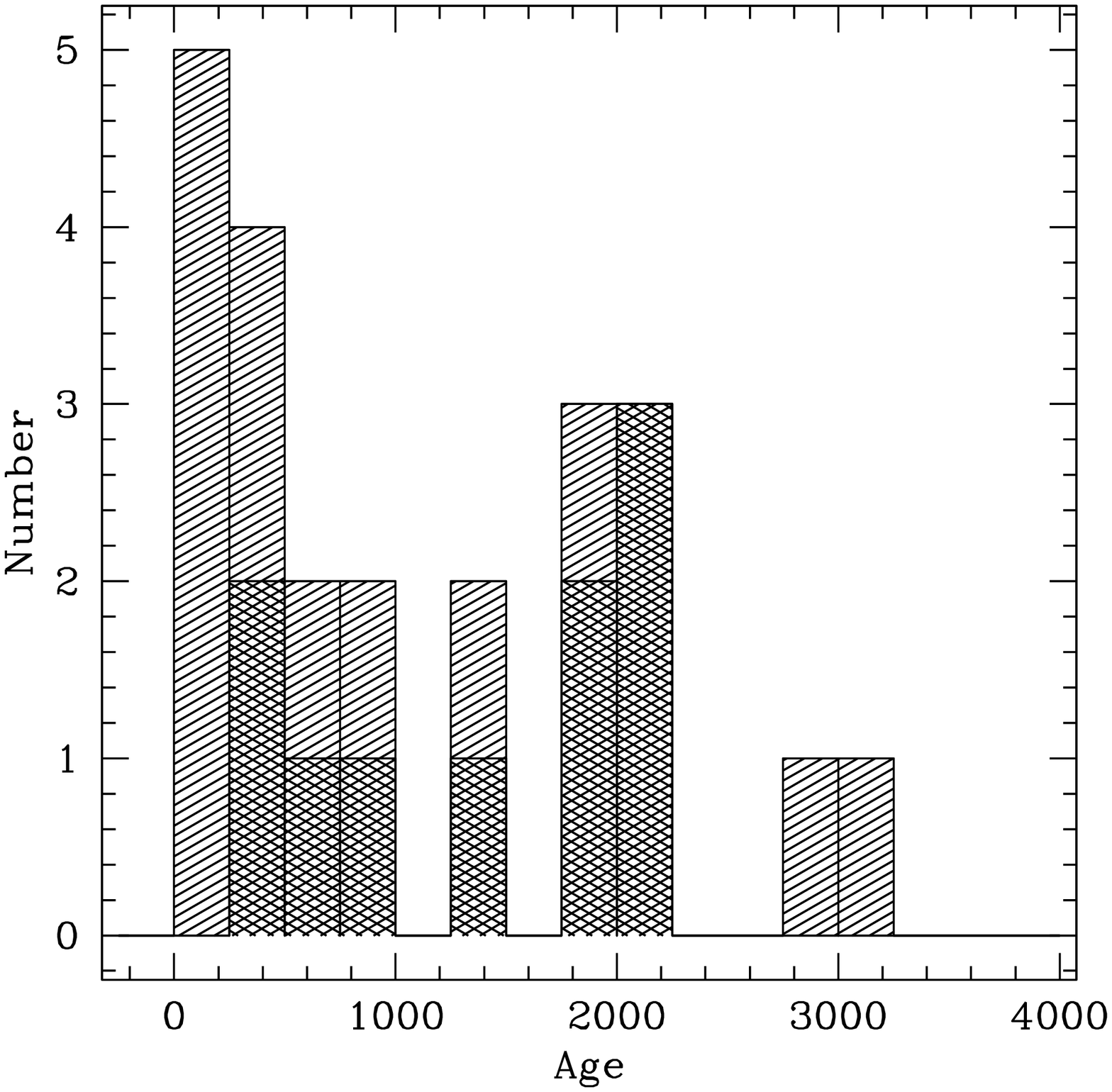}
\caption{Histogram of CSO ages and limits from this paper and those
  cited in Polatidis et al. 2003.  Bars for lower limits have
  cross-hatches while age measurements have single hatching.}
\end{figure}
\clearpage

%Figure 7 -- Total Fluxes
\begin{figure}
\figurenum{7}
\vspace{20.4cm}
\includegraphics{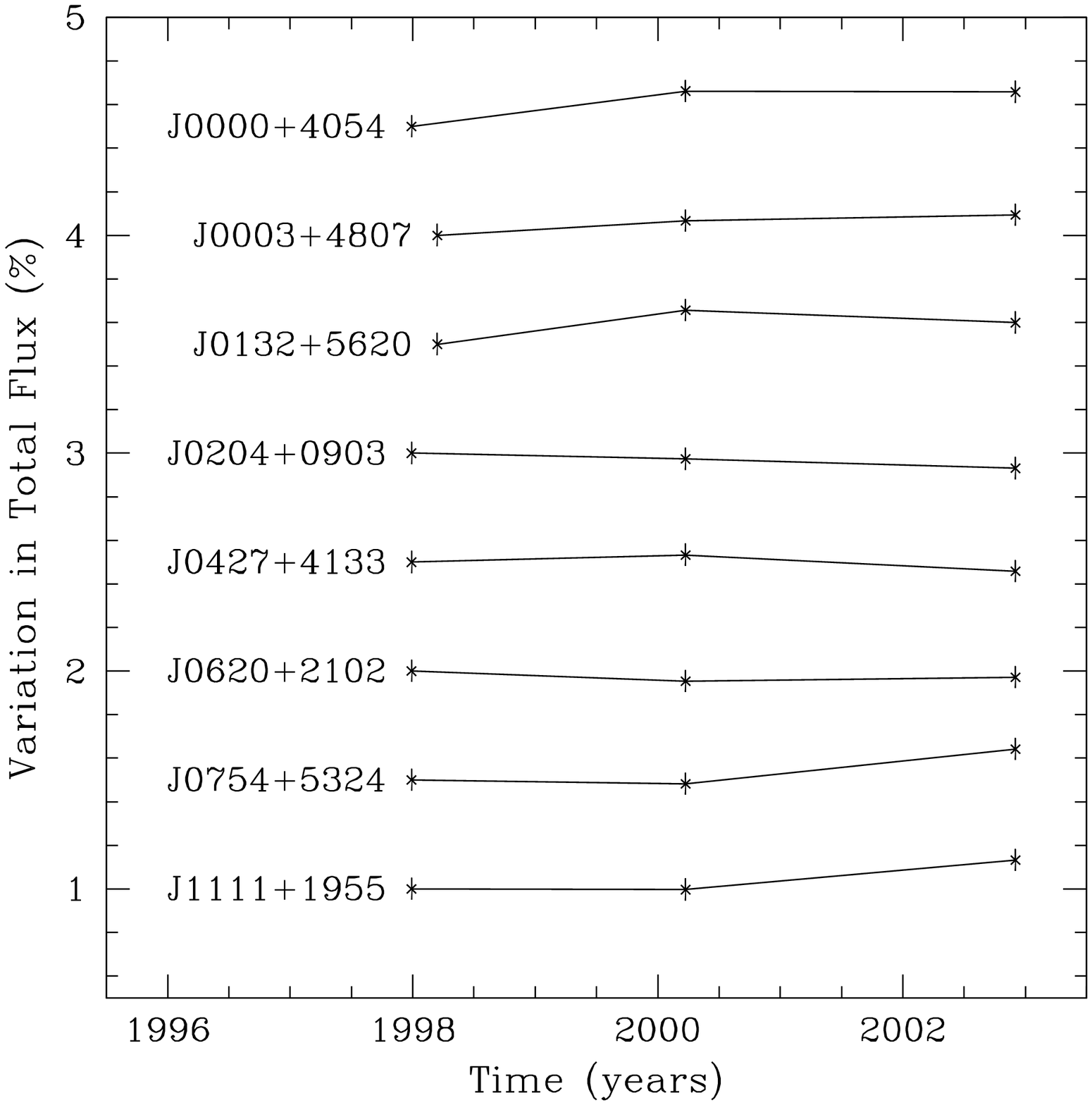}
\includegraphics{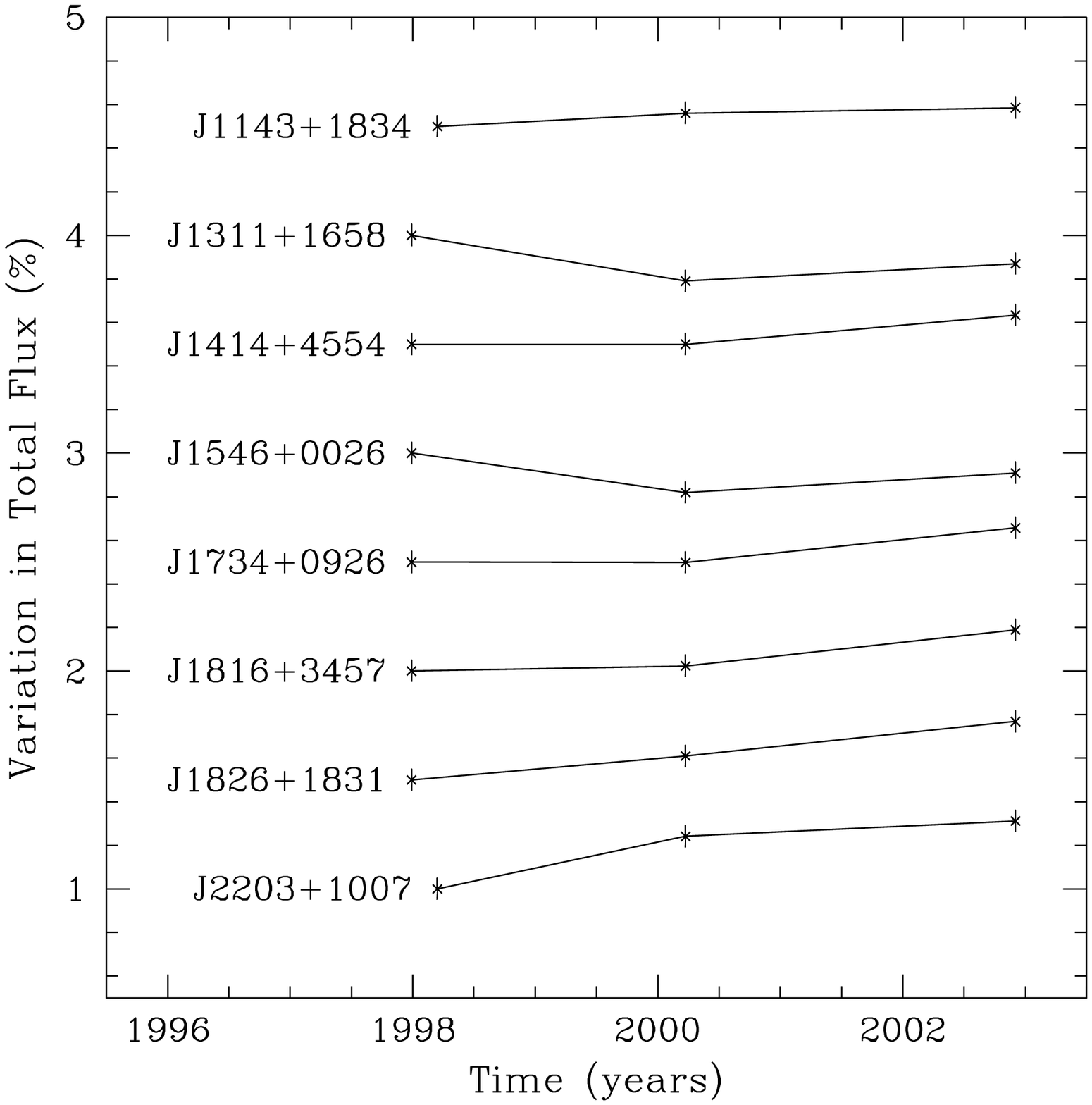}
\caption{Changes in the total flux densities of CSOs and candidates at 8.4~GHz
  over three epochs.  Each source has been offset by a multiple of 50\% in 
order to improve legibility.}
\end{figure}
\clearpage  

\begin{deluxetable}{llllcll}
\tabletypesize{\footnotesize}
\tablewidth{0pt}
\tablecolumns{7}
\tablecaption{VCS Sources in the COINS Sample\label{tab1}}
\tablehead{\colhead{Source} & \colhead{Alternate} &\colhead{} & \colhead{} &
  \colhead{} & \colhead{} & \colhead{} \\
  \colhead{Name} & \colhead{Name} & \colhead{RA} &
  \colhead{Dec} &\colhead{ID} & \colhead{$M_v$} & \colhead{$z$} \\
  \colhead{(1)} & \colhead{(2)} & \colhead{(3)} &
  \colhead{(4)} & \colhead{(5)} &\colhead{(6)} &  \colhead{(7)}}
\startdata
J0000{\tt +}4054 & 4C 40.52 & 00 00 53.081551 & $+$40 54 01.79335  &
G& 21.4 & ... \\
J0003{\tt +}4807 &  &00 03 46.0413    & $+$48 07 04.134$^{(8)}$ & ...&
... & ... \\
J0132{\tt +}5620 &  & 01 32 20.447300 & $+$56 20 40.37012  & ...&
... & ... \\
J0204{\tt +}0903 &  & 02 04 34.759128 & $+$09 03 49.25951  & ...&
... & ... \\
J0332{\tt +}6753 &  & 03 32 59.5241   & $+$67 53 03.860$^{(8)}$ & ...&
... & ... \\
J0427{\tt +}4133 &  &04 27 46.045557 & $+$41 33 01.09988  & ...& ... &
... \\
J0518{\tt +}4730 &  &05 18 12.089801 & $+$47 30 55.52822  & ...& ... &
... \\
J0620{\tt +}2102 &  &06 20 19.528205 & $+$21 02 29.54736  &
...&... &...\\
J0754{\tt +}5324 &  & 07 54 15.2177  & $+$53 24 56.450$^{(8)}$ & ...&
... & ...\\
J1111{\tt +}1955 &  &11 11 20.065804 & $+$19 55 36.00111 & G& 18.5
&0.299\\
J1143{\tt +}1834 &  &11 43 26.069622 & $+$18 34 38.36170 & ...&
...&...\\
J1311{\tt +}1417 &  &13 11 07.824347 & $+$14 17 46.64778 &
QSO&19.5&1.952\\
J1311{\tt +}1658 &  &13 11 23.820125 & $+$16 58 44.18917
&...&...&...\\
J1414{\tt +}4554 &  &14 14 14.853505 & $+$45 54 48.65441 & G & 19.9 &
0.190\\
J1415{\tt +}1320 & PKS~1413$+$135 &14 15 58.817480 & $+$13 20 23.71257  &
QSO & 19.88 & 0.250\\
J1546{\tt +}0026 &  &15 46 09.531469 & $+$00 26 24.61396 & G & 18.45
&0.55\\
J1734{\tt +}0926 &  &17 34 58.376995 & $+$09 26 58.25998 & G & 20.7 &
0.61\\
J1816{\tt +}3457 &  &18 16 23.900825 & $+$34 57 45.74809 & G & 18.7 &
0.245\\
J1826{\tt +}1831 &  &18 26 17.710882 & $+$18 31 52.88973 & ... &
... &...\\
J2203{\tt +}1007 &  &22 03 30.952632 & $+$10 07 42.58629 & ... &
22.53&...\\
J2245{\tt +}0324 &  &22 45 28.284769 & $+$03 24 08.86418 & QSO & 19.0&
1.34\\
\enddata
\tablenotetext{*}{
Notes - (1) J2000 source name; (2) Alternate name; (3) Right
ascension and (4) Declination in J2000 coordinates from the VLBA
Calibrator Survey by Beasley et al. 2002, unless otherwise noted; (5)
Optical host galaxy identification; (6) Optical magnitude; (7)
Redshift (see discussion of individual sources for references); (8)
From Jodrell Bank/VLA Survey with positional accuracy of 20 mas.}
\end{deluxetable}
\clearpage

\begin{deluxetable}{llrrrccc}
\tabletypesize{\footnotesize}
\tablewidth{0pt}
\tablecolumns{8}
\tablecaption{CSO and Candidate Image Parameters\label{tab2}}
\tablehead{
  \colhead{} & \colhead{} & \colhead{} & \colhead{} & 
  \colhead{Peak Flux} & 
  \colhead{rms} & 
  \colhead{Lowest} &
  \colhead{}  \\
  \colhead{Source} & 
  \colhead{Beam} & 
  \colhead{$\theta$} & 
  \colhead{Total Flux} &
  \colhead{(mJy} & 
  \colhead{(mJy} & 
  \colhead{Contour} & 
  \colhead{}  \\
  \colhead{Name} & 
  \colhead{(mas)} & 
  \colhead{($^{\circ}$)} & 
  \colhead{(mJy)} &
  \colhead{beam$^{-1}$)} & 
  \colhead{beam$^{-1}$)} & 
  \colhead{(mJy beam$^{-1}$)} &
  \colhead{Status}
  }
\startdata
J0000{\tt +}4054 & 1.48$\times$1.20 & $-$17.1 & 301.2& 42.0 & 0.1 & 0.3 & 
CSO \\
J0003{\tt +}4807 & 1.48$\times$1.20 & $-$36.4 & 79.6& 26.1 & 0.1 & 0.3 & 
CSO \\
J0132{\tt +}5620 & 1.63$\times$0.98 & $-$11.7 & 326.4& 229.1 & 0.1 & 0.5 
& CAND \\
J0204{\tt +}0903 & 2.15$\times$0.96 & $-$2.1 & 366.3& 64.1 & 0.3 & 0.9 & CSO \\
J0427{\tt +}4133 & 1.73$\times$1.14 & 18.2 & 665.1& 515.0 & 0.1 & 0.4 & CSO\\
J0620{\tt +}2102 & 2.19$\times$1.05 & $-$15.0 & 267.3& 127.0 & 0.1 & 0.4 &CSO\\
J0754{\tt +}5324 & 1.83$\times$1.00 & $-$17.0 & 111.3& 26.4 & 0.1 & 0.3 & CSO\\
J1111{\tt +}1955 & 2.14$\times$0.94 & $-$3.1 & 324.4& 62.5 & 0.2 & 0.7 & CSO\\
J1143{\tt +}1834 & 2.13$\times$1.03 & 0.8 & 250.6& 118.1 & 0.1 & 0.3 & CSO\\
J1311{\tt +}1658 & 2.03$\times$0.98 & 2.7 & 198.8& 74.5 & 0.3 & 0.7 & CAND\\
J1414{\tt +}4554 & 1.73$\times$1.10 & 23.1 & 120.4& 15.8 & 0.1 & 0.3 & CSO\\
J1415{\tt +}1320 & 2.15$\times$0.95 & $-$7.1 & 609.7& 376.6 & 0.1 & 0.4 & CSO\\
J1546{\tt +}0026 & 2.22$\times$0.93 & $-$2.0 & 542.7& 177.1 & 0.1 & 0.3 & CSO\\
J1734{\tt +}0926 & 2.17$\times$0.94 & $-$3.8 & 496.8& 182.3 & 0.1 & 0.3 & CSO\\
J1816{\tt +}3457 & 1.84$\times$0.96 & $-$3.4 & 217.0& 29.8 & 0.1 & 0.3 & CSO\\
J1826{\tt +}1831 & 2.03$\times$0.97 & $-$2.9 & 313.7& 48.5 & 0.1 & 0.4 & CSO\\
J2203{\tt +}1007 & 2.15$\times$0.94 & $-$2.2 & 210.7& 99.7 & 0.1 & 0.3 & CSO\\
\enddata
\end{deluxetable}
\clearpage

\begin{deluxetable}{llrrccc}
\tabletypesize{\footnotesize}
\tablewidth{0pt}
\tablecolumns{7}
\tablecaption{Core-Jet Image Parameters\label{tab3}}
\tablehead{\colhead{} & \colhead{} &\colhead{} & \colhead{Peak Flux} &
  \colhead{rms} & \colhead{Lowest} & \colhead{} \\
  \colhead{Source} & \colhead{Beam} & \colhead{$\theta$} &
  \colhead{(mJy} &\colhead{(mJy} & \colhead{Contour} & \colhead{} \\
  \colhead{Name} & \colhead{(mas)} & \colhead{} &
  \colhead{beam$^{-1}$)} & \colhead{beam$^{-1}$)} &\colhead{(mJy)} &
  \colhead{}}
\startdata
J0332{\tt +}6753 & 1.54$\times$1.00 & 14.0 & 13.1 & 0.1 & 0.3 \\
J0518{\tt +}4730 & 1.76$\times$1.06 & 22.0 & 131.3 & 0.1 & 0.3  \\
J1311{\tt +}1417 & 2.16$\times$1.00 & 0.4 & 101.8 & 0.1 & 0.3  \\
J2245{\tt +}0324 & 2.25$\times$0.99 & $-$2.2 & 190.0 & 0.1 & 0.4  \\
\enddata
\end{deluxetable}
\clearpage

\begin{deluxetable}{lcrrrrrcccr}
\tabletypesize{\footnotesize}
\tablewidth{0pt}
\tablecolumns{11}
\tablecaption{CSO and Candidate Component Modelfits, Polarization, and
  Motions\label{tab4}}
\tablehead{\colhead{ } & \colhead{ }&\colhead{$S_{epoch1}$} &
  \colhead{$S_{epoch 2}$} & \colhead{$S_{epoch 3}$} &\colhead{Core}& \colhead{ } &
  \colhead{$P_{epoch 2}$} & \colhead{$\mu$} & \colhead{$v$}
  &\colhead{P.A.} \\
  \colhead{Source} & \colhead{Comp.}&\colhead{(mJy)} &
  \colhead{(mJy)} & \colhead{(mJy)} &\colhead{Fraction}& 
  \colhead{$\alpha^{8.4}_{15}$} &
  \colhead{(mJy)} & \colhead{(mas yr$^{-1}$)} & \colhead{$(c)$}
  &\colhead{(deg)} \\
  \colhead{(1)} & \colhead{(2)}&\colhead{(3)} &
  \colhead{(4)} & \colhead{(5)} & \colhead{(6)} &
  \colhead{(7)} & \colhead{(8)} & \colhead{(9)}
  &\colhead{(10)} &\colhead{(11)}}
\startdata
J0000{\tt +}4054 &A& 11 & 14 & 16 & ...&... & $<$0.3 & $<$0.144
 & n/a & ... \\
     &B1& 8 & 13 & 14 & ...& $-$1.8 & ... & ... & ... & ... \\
     &B2& 28 & 26 & 27 & ...& $-$1.1 & ... & ... & ... & ... \\
     &C& 172 & 186 & 184 & ...& $-$1.2 & 0.9  & reference & ... & ... \\
J0003{\tt +}4807 &A& 36 & 38 & 39 &...& n/a$^{12}$ & $<$0.3 &
 $<$0.014 & n/a & ...\\
     &B& 4 & 3 & 3 & 4\% & ... & ...& reference & ... & ... \\
     &C1& 15 & 13 & 12 &...& ...& ... & ... & ... & ... \\
     &C2& 26 & 26 & 27 & ...&...& ... & ... & ... & ... \\
J0132{\tt +}5620 &A& 34 & 37 & 35 &...& $-$25 & $<$0.3 & $<$0.014
 & n/a & ... \\
     &B& 8 & 8 & 10 & ...& ... & ... & ... & ... & ...  \\
     &C& 12 & 11 & 9 & ...& $-$1.0 & ... &  $<$0.054 & n/a & ... \\
     &D1& 34 & 58 & 62 & ...& $-$1.5 & ... & 0.015$\pm$0.007 & n/a &
$-$90\\
     &D2& 205 & 209 & 194 & ...& $-$1.5 & ... & reference & ... &
        ...\\
J0204{\tt +}0903 &A& 129 & 120 & 120 &...& n/a$^{12}$ & $<$0.9  &
 $<$0.024 & n/a & ...\\
     &B& 81 & 74 & 71 &...& ... & ... & $<$0.014 & n/a & ... \\
     &C& 67 & 72 & 58 & 20\% &... & ... & reference & ... & ...\\
     &D& 71 & 68 & 66 & ...& ... & ... & 0.070$\pm$0.011 & n/a &
       $-$90\\
J0427{\tt +}4133 &A& 25 & 28 & 68 &...& n/a$^{12}$ & $<$0.3 &  ... & ... &
...\\
     &B& 561 & 571 & 490 & 86\% & ... & ... & reference & ... & ...\\
     &C& 66 & 65 & 51 & ...& ... & ... & 0.060$\pm$0.013 & n/a &
      $-$63 \\
J0620{\tt +}2102 &A& 91 & 88 & 88 &...& $-$1.0 & $<$0.3 &
 $<$0.013 & n/a & ...\\
     &B& 167 &156 &150 & ...& $-$1.0 & ... & reference &
        ... &... \\
J0754{\tt +}5324 &A& 33& 36& 40 & ...& n/a$^{12}$& $<$0.3 &
 $<$0.009 & n/a & ...\\
     &B& 11& 11& 13 & ...& ... & ... & 0.060$\pm$0.026 & n/a & $-$51\\
     &C& 38& 39& 42 & ...& ...& ...& reference & ...& ...\\
J1111{\tt +}1955 &A& 78& 76& 86& ...& $-$0.9 & $<$0.6 & reference
 & ... &...\\
     &B1& 96& 98& 111& ...& $-$0.8 & ... & $<$0.010 & $<$0.14
       &...\\
     &B2& 77& 77& 80& ...& $-$0.8 & ... & $<$0.013 &
        $<$0.19 & ...\\
J1143{\tt +}1834 &A&106& 99& 97& ...& $-$0.8 & $<$0.3 & $<$0.010 &
 n/a & ... \\
     &B& 124& 130& 127& ...& $-$0.8 & ... & reference& ...&...\\
J1311{\tt +}1658 &A& 130& 130& 170& ...& ... & $<$0.9 & ...&...&...\\
     &B1& 107& 52& 58& ...& $-$0.2 & ... & ... & ... & ...\\
     &B2& 111& 102& 110& ...& $-$0.2 & ... & reference & ... & ...\\
     &C& 7& 9& 9& ...& $-$0.4 & ... & ... & ...&...\\
     &D& 18& 18& 19& ...& ... & ... & ... & ... & ...\\
J1414{\tt +}4554 &A& 63& 64& 72& ...& n/a$^{12}$ & $<$0.3& reference &
  ... & ...\\
     &B& 46& 51& 54& ...& ... & ... & $<$0.014 & $<$0.14 & ...\\
J1546{\tt +}0026 &A& 196& 213& 236& ...& n/a$^{12}$& $<$0.3& ...& ...&
  ...\\
     &B& 347& 217& 231& 40\% &...& ...& reference & ...& ...\\
     &C& 118& 110& 128& ...& ...& ...& 0.054$\pm$0.008 & 1.10$\pm$0.17 &
  $-$90\\
J1734{\tt +}0926 &A& 114& 119& 142& ...& n/a$^{12}$& $<$0.3& $<$0.008 &
  $<$0.18 & ...\\
     &B& 191& 183& 204& ...& ...& ...& reference &...&...\\
J1816{\tt +}3457 &A& 137& 135& 148& ...& n/a$^{12}$& $<$0.3& reference &
  ... & ...\\
     &B& 58& 59& 65& ...& ...& ...& 0.036$\pm$0.009 & 0.40$\pm$0.11 &
90\\
J1826{\tt +}1831 &A& 14& 16& 18& ...& n/a$^{12}$& $<$0.3& ...& ...&
  ...\\
     &B& 80& 90& 70& 29\% &...& ...& reference& ...& ...\\
     &C& 38& 34& 34& ...& ...& 3.0& 0.037$\pm$0.009& n/a& $-$90\\
     &D& 109& 100& 112& ...& ...& ...& 0.013$\pm$0.006& n/a& $-$90\\
J2203{\tt +}1007 &A1& 78& 82& 83& ...& $-$0.8& $<$0.3& reference& ...&
  ...\\
     &A2& 44& 63& 64& ...& $-$0.8& ...& 0.020$\pm$0.008& n/a& $-$90\\
     &B& 10& 12& 15& ...& $-$0.8& ...& ... & ...& ...\\
     &C& 44& 53& 56& ...& $-$1.3& ...& $<$0.011& n/a& ...\\
\enddata
\tablenotetext{*}{
Notes - (1) Source name; (2) Component (see Figure 1); (3) Integrated
flux density of Gaussian model component in 1997.990 or 1998.201; (4)
Integrated flux density of Gaussian model component in 2000.227; (5)
Integrated
flux density of Gaussian model component in 2002.919; (6) Percentage of total 
flux that is attributed to the core in 2000.227; (7)Spectral
index between 8.4~GHz
(2000.227) and 15~GHz (2001.008); (8) Polarized intensity in
2000.227, or 3$\sigma$ limit; (9) Relative proper motion;
(10) Relative projected velocity (if $z$ available); (11) Projected
direction of velocity; (12) 15~GHz not available for these sources,
see Peck \& Taylor 2000.  Assume 5\% error for flux densities.}
\end{deluxetable}
\clearpage

\begin{deluxetable}{lcrrrrccc}
\tabletypesize{\footnotesize}
\tablewidth{0pt}
\tablecolumns{9}
\tablecaption{J1415+1320 Component Modelfits, Polarization, and
  Motions\label{tab5}}
\tablehead{\colhead{ }&\colhead{$S_{1994}$} &
  \colhead{$S_{1995}$} & \colhead{$S_{1998}$}
  &\colhead{$S_{2000}$} & 
  \colhead{$P_{epoch 2}$} & \colhead{$\mu$} & \colhead{$v$}
  &\colhead{P.A.} \\
   \colhead{Comp.}&\colhead{(mJy)} & \colhead{(mJy)}&
  \colhead{(mJy)} & \colhead{(mJy)}  &
  \colhead{(mJy)} & \colhead{(mas yr$^{-1})$} & \colhead{$(c)$}
  &\colhead{(deg)} \\
  \colhead{(1)} & \colhead{(2)}&\colhead{(3)} &
  \colhead{(4)} & \colhead{(5)} & \colhead{(6)} &
  \colhead{(7)} & \colhead{(8)} & \colhead{(9)}}
\startdata
A& 20& 27& 25& 29& $<$0.3&...&...&...\\
B1& 32& 40& 40& 45& ...& ...& ...& ...\\
B2& 6& 9& 9& 10& ...& 0.060$\pm$0.024& 0.80$\pm$0.30& 90\\
C& 878& 1064& 1266& 435&  ...& reference& ...&...\\
C3& 70& 97& 64& 48& ...& 0.087$\pm$0.009& 1.10$\pm$0.11& $-$102\\
C2& 26& 30& 37& 36& ...& ...&...&...\\
Total& 1032 & 1267 & 1441 & 603 & ...& ...& ...& ...\\
\enddata
\tablenotetext{*}{
Notes - (1) Component (See Figure 1); (2) Integrated
flux density of Gaussian model component in 1994.519; (3) Integrated
flux density of Gaussian model component in 1995.535; (4) Integrated
flux density of Gaussian model component in 1998.201; (5) Integrated
flux density of Gaussian model component in 2000.227; (6) Polarized
intensity in 2000.227, or 3$\sigma$ limit; (7) Relative proper
motion; (8) Relative projected velocity ($z$=0.25); (9) Projected
direction of velocity.  No spectral index map is available (see Peck
\& Taylor 2000).  1994 and 1995 epochs from Perlman et al 1996.  2002.919 could not be properly modelfit.  Assume 5\% error for flux densities.}
\end{deluxetable}
\clearpage

\begin{deluxetable}{lllccc}
\tabletypesize{\footnotesize}
\tablewidth{0pt}
\tablecolumns{6}
\tablecaption{CSO Kinematic Component Ages Where Redshift is
  Not Available\label{tab6}}
\tablehead{\colhead{}&\colhead{}&\colhead{}& \colhead{$\mu$} &\colhead{Separation}
  &\colhead{Kin. Age}\\
  \colhead{Source} &\colhead{Comps.} &\colhead{Type}&\colhead{(mas yr$^{-1}$)}
  &\colhead{(mas)} &\colhead{(yr)}\\
   \colhead{(1)}& \colhead{(2)}&\colhead{(3)} &
  \colhead{(4)} & \colhead{(5)} & \colhead{(6)}}
\startdata
J0000{\tt +}4054 &C to A& h to h& $<$0.144& 40.33 & $>$280\\
J0003{\tt +}4807 &B to A& c to h& $<$0.014& 4.78 & $>$340\\
J0204{\tt +}0903 &C to D& c to h& 0.070$\pm$0.011& 18.27
                   &240$\pm$36\\
J0427{\tt +}4133 &B to C& c to h?& 0.060$\pm$0.013& 1.33&
                   20$\pm$4\\
J0620{\tt +}2102 &B to A& h to h& $<$0.013& 26.74& $>$2060\\
J0754{\tt +}5324 &C to A& h to h& $<$0.009& 20.01& $>$2220\\
J1143{\tt +}1834 &B to A& h to h& $<$0.010& 6.87& $>$690\\
J1826{\tt +}1831 &B to C& c to j& 0.037$\pm$0.009& 14.18&
                   380$\pm$93\\
                 &B to D& c to h& 0.013$\pm$0.006& 41.87& 3000$\pm$1490\\
J2203{\tt +}1007 &A1 to C& h to h& $<$0.011& 10.30& $>$940\\
\enddata
\tablenotetext{*}{
Notes - (1) Source name; (2) Components (see figures); (3) Component
types (h - hot spot, c - core, j - jet); (4) Relative proper motion;
(5) Distance between components (2000.227); (6) Kinematic age estimate.}
\end{deluxetable}
\clearpage

\begin{deluxetable}{lllccc}
\tabletypesize{\footnotesize}
\tablewidth{0pt}
\tablecolumns{6}
\tablecaption{CSO Kinematic Component Ages Where Redshift is
  Available\label{tab7}}
\tablehead{\colhead{}&\colhead{}&\colhead{}& \colhead{$v$} &\colhead{Separation}
  &\colhead{Kin. Age}\\
  \colhead{Source} &\colhead{Comps.} &\colhead{Type}&\colhead{(c)} &\colhead{pc}
  &\colhead{(yr)}\\
   \colhead{(1)}& \colhead{(2)}&\colhead{(3)} &
  \colhead{(4)} & \colhead{(5)} &\colhead{(6)}}
\startdata
J1111{\tt +}1955 &A to B2& h to h& $<$0.14& 71.05& $>$1620\\
                 &A to B1& h to h& $<$0.19& 77.88&
  $>$1360\\
J1414{\tt +}4554 &A to B& h to h& $<$0.14& 88.93& $>$2030\\
J1415{\tt +}1320 &C to B2& c to j& 0.800$\pm$0.30& 31.12&
130$\pm$47\\
                 &C to C3& c to j& 1.10$\pm$0.11& 7.63& 22$\pm$2\\
J1734{\tt +}0926 &B to A& h to h& $<$0.18& 95.50 & $>$1780\\
\enddata
\tablenotetext{*}{
Notes - (1) Source name; (2) Components (see figures); (3) Component
types (h - hot spot, c - core, j - jet); (4) Relative proper motion;
(5) Distance between components (2000.227); (6) Kinematic age
estimate.}
\end{deluxetable}
\clearpage

\begin{deluxetable}{lcrrrrrccc}
\tabletypesize{\footnotesize}
\tablecolumns{10}
\tablecaption{Core-Jet Component Modelfits, Polarization, and
  Motions\label{tab8}}
\tablehead{\colhead{ } & \colhead{ }&\colhead{$S_{epoch1}$} &
  \colhead{$S_{epoch 2}$} & \colhead{$S_{epoch 3}$} & \colhead{ } &
  \colhead{$P_{epoch 2}$} & \colhead{$\mu$} & \colhead{$v$}
  &\colhead{P.A.} \\
  \colhead{Source} & \colhead{Comp.}&\colhead{(mJy)} &
  \colhead{(mJy)} & \colhead{(mJy)} & \colhead{$\alpha^{8.4}_{15}$} &
  \colhead{(mJy)} & \colhead{(mas yr$^{-1})$} & \colhead{$(c)$}
  &\colhead{(deg)} \\
  \colhead{(1)} & \colhead{(2)}&\colhead{(3)} &
  \colhead{(4)} & \colhead{(5)} & \colhead{(6)} &
  \colhead{(7)} & \colhead{(8)} & \colhead{(9)}
  &\colhead{(10)}}
\startdata
J0332{\tt +}6753 &A& 50& 51& 53& $-$0.5 & 1.7 & $<$0.022 &n/a& ...\\
     &B& 21& 22& 25& $-$0.8& $<$0.3& $<$0.014& n/a& ...\\
     &C& 80& 130& 70& $-$0.1& ...& reference & ...&...\\
J0518{\tt +}4730 &A& 8& 13& 7& ...& $<$0.3& reference& ...& ...\\
     &B& 50& 51& 53& $-$0.6& ...& 0.026$\pm$0.009& n/a& $-$85.3\\
     &C& 21& 22& 25& $-$0.8& 0.8& 0.018$\pm$0.009& n/a& $-$159.4\\
J1311{\tt +}1417 &A& 131& 127& 122& $-$1.0& 4.4 & ... &... &... \\
     &B& 106& 119& 131& $-$1.0& 9.0 & ...& ...& ...\\
     &C& 13& 22& 17& $-$0.3& $<$0.3& reference& ...& ...\\
J2245{\tt +}0324 &A& 111& 121& 118& $-$0.5& 3.1 & 0.044$\pm$0.007 &
  1.20$\pm$0.19& 90.0\\
     &B& 293& 239& 244& $-$0.2& $<$0.3& reference& ...& ...\\
\enddata
\vspace{0.5cm}
\tablenotetext{*}{Notes - (1) Source name; (2) Component (see Figure 2); (3) Integrated
flux density of Gaussian model component in 1997.990 or 1998.201; (4)
Integrated flux density of Gaussian model component in 2000.227; (5)
Integrated
flux density of Gaussian model component in 2002.919; (6) Spectral
index between 8.4~GHz
(2000.227) and 15~GHz (2001.008); (7) Polarized intensity in
2000.227, or 3$\sigma$ limit; (8) Relative proper motion;
(9) Relative projected velocity (if $z$ available); (10) Projected
direction of velocity.  Assume 5\% error for flux densities.}
\end{deluxetable}
\clearpage

\end{document}